\documentclass[11pt]{article}
\usepackage{mathpazo}
\usepackage{natbib}
\usepackage{amsmath}
\usepackage{amsthm}
\usepackage{multicol}
\usepackage{amssymb}
\usepackage[]{graphicx}
\usepackage{longtable}
\usepackage{lscape}
\usepackage{epstopdf}
\usepackage{rotating}
\usepackage{setspace}
\usepackage{url}

\bibpunct{(}{)}{;}{a}{}{,}
\bibdata{netcaus}
\makeatletter
\let\ps@normal\ps@plain
\def\ps@plain{}
\makeatother

\title{\bfseries\Large{The ``Unfriending'' Problem}\\
\bfseries\large{The Consequences of Homophily in Friendship Retention \\ for Causal Estimates of Social Influence}
}
\author{Hans Noel\\ Assistant Professor of Government\\ Georgetown University\\   hcn4@georgetown.edu\and
Brendan Nyhan\\
RWJ Scholar in Health Policy Research\\
University of Michigan\\
bnyhan@umich.edu
}

\begin{document}
\large
\DeclareGraphicsExtensions{.eps}
\maketitle

\begin{abstract} 
\noindent An increasing number of scholars are using longitudinal social network data to try to obtain estimates of peer or social influence effects. These data may provide additional statistical leverage, but they can introduce new inferential problems. In particular, while the confounding effects of homophily in friendship formation are widely appreciated, homophily in friendship \emph{retention} may also confound causal estimates of social influence in longitudinal network data. We provide evidence for this claim in a Monte Carlo analysis of the statistical model used by Christakis, Fowler, and their colleagues in numerous articles estimating ``contagion'' effects in social networks. Our results indicate that homophily in friendship retention induces significant upward bias and decreased coverage levels in the Christakis and Fowler model if there is non-negligible friendship attrition over time.
\end{abstract}
\vspace{8pt}

\thanks{\noindent \footnotesize{We thank Nicholas Christakis and James Fowler for generously sharing the simulation code adapted in this paper and for helpful comments and clarifications. We are also grateful to Megan Andrew, Jason Fletcher, Rob Franzese, Michael Heaney, Jonathan Ladd, Russell Lyons, Aya Kachi, Walter Mebane, Jacob Montgomery, David Nickerson, Edward Norton, Phil Paolino, Fabio Rojas, Betsy Sinclair and two anonymous reviewers for helpful suggestions and feedback.}}

\newpage
\doublespacing

\pagestyle{normal}
\pagenumbering{arabic}
\onehalfspacing
\normalsize

\section{Introduction}
Until recently, scholars have given relatively little attention to the influence of personal relationships on human behavior, instead studying people largely as atomistic individuals ripped from the social context in which they live. Thankfully, this impoverished approach has started to give way to an interdisciplinary movement seeking to understand the influence of social networks in domains ranging from health to politics. Results from cases in which peers were randomly or quasi-randomly assigned such as college roommates have provided credible evidence of such effects (e.g., \citealt{sacerdote01,zimmerman03,carrell11}; for a review, see \citealt{kremer08}).\footnote{Related experimental studies by \citet{nickerson08} and \citet{fowler10} provide suggestive evidence that such effects can extend two or more degrees, though the applicability of their results to non-experimental contexts is unclear.} 

However, when random assignment of peers is not feasible, researchers must use observational data, which creates serious inferential problems \citep{manski93}. In particular, peers may behave similarly as a result of ``correlated effects'' such as common environmental shocks or shared characteristics rather than social influence. Given the likelihood that peers will be similar on a range of characteristics due to homophily \citep{mcpherson01}, distinguishing between homophily and peer effects has proven to be a very difficult challenge.\footnote{For a review of the literature, see \citet{soetevent06}.}

Many scholars have therefore turned to use longitudinal network data to try to separate homophily from social influence effects. In principle, observing dyads over multiple periods seems as though it could help separate homophily in tie formation from subsequent peer influence. However, homophily may also affect whether social ties are \emph{maintained} over time, confounding estimates of peer influence effects. We call this the ``unfriending'' problem in honor of the Facebook practice of removing a person from one's list of friends on the online social network site.

We illustrate the potential inferential consequences of this problem below in an analysis of the statistical model used to estimate ``contagion'' effects in a series of widely publicized studies by Christakis, Fowler, and their colleagues (hereafter CF). Our Monte Carlo simulations, which are adapted from those of CF, indicate that their model's estimates of social influence effects are unbiased and have accurate coverage levels when homophily in friendship retention is not present. However, when non-negligible attrition is present, estimates from the CF model show substantial upward bias and decreased coverage levels as homophily in friendship retention increases. In short, the ``unfriending'' problem can create spurious evidence of social influence when none exists.

\section{Leveraging dynamic networks: A solution?}

The CF studies, which use longitudinal social network data from the Framingham Heart Study (FHS) and National Longitudinal Study of Adolescent Health (Add Health), make strong claims about the effects of one's friends\footnote{The studies typically also estimate social influence effects among family members; we do not consider the validity of those estimates here.} on a wide range of dependent variables: obesity \citep{christakis07,fowlerjhe08}, smoking \citep{christakis08}, happiness \citep{fowler08}, loneliness \citep{cacioppo09}, depression \citep{rosenquist10a}, alcohol consumption \citep{rosenquist10b}, sleep loss \citep{mednick10}, and divorce \citep{mcdermottnd}. Each CF paper uses the same approach, estimating versions of the following model for ego $i$ and alter $j$ observed at times $t_0$ and $t_1$:
\begin{equation}
Y_\mathrm{i,t_1}=f(Y_\mathrm{i,t_0}, Y_\mathrm{j,t_0}, Y_\mathrm{j,t_1}, \text{controls}) 
\end{equation}
These models are typically estimated using generalized estimating equations (GEE) with an independent correlation structure to account for repeated observations of the same ego (specifically, those who name or are named by more than one friend) and dyad (those who name each other and are thus included twice, one each as the ego and once as the alter). The functional form of the model varies depending on the distribution of the dependent variable (logistic regression if the dependent variable is binary; linear regression if it is continuous or quasi-continuous).\footnote{For other approaches to obtaining causal estimates of peer effects in longitudinal or repeatedly sampled network data, see \citet{anagnostopoulos08}, \citet{bramoulle09}, \citet{aral09}, and \citet{lazer10}.} 

CF argue that this specification controls for initial homophily (i.e., the likely similarity between $Y_{i,t_0}$ and $Y_{j,t_0}$), allowing us to identify the causal effect of \emph{changes} in the alter's trait from $t_0$ to $t_1$ by estimating the effect of $Y_{j,t_1}$ controlling for $Y_{j,t_0}$. In \citet{christakis07}, they write that including alters' lagged obesity as a covariate ``controlled for homophily'' (373). In later work, the language is somewhat more hedged---for instance, they write in \citet[2251]{christakis08} that a lagged measure of alter smoking ``\emph{helped} to account for homophily'' (our emphasis)---but the suggestion that the coefficient for $Y_{j,t_1}$ is a causal estimate of peer effects remains. \citet{cfbook} expands on these claims, stating that observed clustering at up to three degrees of separation reflects ``Three Degrees of Influence''  for happiness (51), obesity (108), and smoking (116) and asserting that we ``now know that obesity is contagious'' (111).

\citet{ccf08b,ccf08a} and \citet{halliday09} question whether CF's model adequately controls for homophily, which has been shown to be significant for weight status \citep{trogdon08,halliday09,valente09}\footnote{\citet{delahaye10} finds homophily in obesity-related behaviors as well. For further explorations of possible social transmission of obesity or weight status, see \citet{anderson09}, \citet{barnes10}, \citet{brown10}, \citet{mcferran10b}, \citet{mcferran10a}, and \citet{campbell11}.}, and suggest that their model may generate spurious inferences (see also \citealt{shalizi11}, \citealt{lyons10}, and \citealt{ellen09}).\footnote{There are other concerns with this statistical model such as possible simultaneity bias and environmental confounding that we do not discuss here.} In response, CF describe Monte Carlo simulation results ``documenting that homophily (ranging from no homophily to complete homophily) does not result in bias in the estimates of induction in this model specification'' \citep[1404]{fowlerjhe08}.

CF's Monte Carlo results, which are presented in \citet{fowlerjhe08} and in a very similar form in \citet{fowler11}, are derived from a stylized model in which a population of individuals with a randomly chosen value on some characteristic of interest form friendships and then influence each other or not (we discuss the procedure in more detail below). CF find that estimates of this influence coefficient are approximately unbiased across varying levels of homophily when the true peer effect is 0 and have a slight downward bias when the true peer effect is 0.1. On this basis, they conclude that ``This simulation evidence suggests that the [Cohen-Cole and Fletcher] assertion that homophily causes us to overestimate the size of the induction effect is false.'' However, as we discuss below, their simulation does not incorporate friendship \emph{attrition} and thus fails to fully account for the effects of homophily.

\section{The ``unfriending'' problem in longitudinal data}\label{unfriending}

Due to the prevalence of cross-sectional data and interest in fixed characteristics such as race and sex, scholars of social networks have tended to think about homophily in relatively static terms and to analyze it as a propensity to form ties with others who share similar characteristics. However, social networks are actually the result of a dynamic process of friendship \emph{formation} and \emph{dissolution}. 

As a result, while relatively few longitudinal network studies have been conducted, most report substantial levels of friendship dissolution between survey waves. For instance, \citet{Mollenhorst09} finds that about half of adult friendships were replaced over the seven years between the two waves of his survey. For the adolescents in Add Health, \citet{moody99} found approximately half of the friends named by respondents during in-school interviews were named again during in-home interviews six to eight months later. Studies of social networks among younger children have found rates of attrition that are even higher still \citep{hallinan87,cairns95}. A partial exception is the FHS data used by CF, which was conducted in a relatively stable community. \citet{omalleynd} report that 82\% of friendships were maintained between waves in FHS, which could be a result of asking for ``close friends'' who could help the researchers contact the participant in the future.\footnote{\label{point96fn} CF report that they treated friendship ties as maintained when a friend as named at $t_{1}$ and $t_{3}$ but tie information was missing at $t_{2}$ (personal communication). Under this definition, 96\% of friendship ties were maintained between waves. However, since missing tie information was often the result of missing an exam altogether, friendship retention in their statistical analyses is likely to be lower.}

When friendship attrition is present, homophily is likely to be a factor. Just as people who are similar are more likely to be friends, friends who are less similar are more likely to \emph{stop} being friends. Most of us have had friends from whom we have grown apart in this way. As we have less in common with those people, we stop spending time with them and eventually fall out of touch. In some cases, one person may deliberately end the relationship as a result of differences in political views, alcohol consumption, or other behaviors or characteristics.

Numerous examples of homophily in tie dissolution have been documented in the sociology literature (see \citealt{burt00} and \citealt{mcpherson01} for reviews). One well-known example is a two-wave study of adolescent friendships by \citet{kandel78}. She describes homophily in friendship retention based on both initial characteristics and subsequent behavior (430): 
\begin{quote}
At time 1, prior to any subsequent change, pairs that will remain stable over time are much more similar in their behaviors and values [marijuana use, educational goals, political views, and delinquency] than the subsequently unstable pairs... At time 2, homophily among former friends is lower than among new friendship pairs or stable pairs.
\end{quote}
She interprets these results as a combination of selection (choosing to become and stay friends with those who are like you) and socialization (acting more like your friends in those relationships you maintain) (433--435):
\begin{quote}
The results support the general conclusion that adolescents coordinate their choices of friends and their behaviors, in particular the use of marijuana, so as to maximize congruency within the friendship pair. If there is a state of unbalance such that the friend's attitude or behavior is incongruent with the adolescent's, the adolescent will either break off the friendship and seek another friend or will keep the friend and modify his own drug behavior.
\end{quote}
Childhood and adolescent friendships have also been found to be more stable when students are more alike by gender \citep{tuma79,degirmencioglu98,moody99}, race \citep{moody99}, grade \citep{moody99}, achievement/competence \citep{tuma79,newcomb99}, and aggression \citep{newcomb99}. 

Similar patterns have been observed among adults. For instance, \citet{kossinets09} find that dyads who are more similar demographically are less likely to break ties in the email network of a large American university (433--434); \citet{popielarz95} show that members of voluntary groups who are dissimilar from other members are more likely to leave; and \citet{burt00} documents homophily in the maintenance and dissolution of bankers' collegial relationships along several dimensions. Most notably, \cite{omalleynd} document homophily in friendship retention within the FHS data used in almost all of CF's studies. They find that differences in BMI, smoking, and measures of body type are significantly related to the dissolution of friendship ties. 

These patterns of homophily in unfriending are potentially a problem for any statistical analysis of peer effects in observational data. Social network scholars have been concerned for some time about the difficult of separating homophily in friendship formation from contextual influences and peer effects. However, homophily in friendship retention is an equally difficult problem. In particular, unfriending is a significant concern for the CF approach. The specification of their generalized estimating equation models requires an ego to name an alter as a friend for two or more consecutive waves (although see footnote \ref{point96fn} above). In this way, they attempt to leverage longitudinal network data to control for past friendship ties. However, what happens when some of the dyads at $t_0$ are no longer friends at $t_1$? Fowler and Christakis argue that including such friendship pairs in their data will bias the results against finding an effect because ``it essentially adds `random' non-friend relationships (i.e., people who are no longer friends) to the pool of friends'' \citep[1401]{fowlerjhe08}. This is a legitimate issue; non-friends presumably can no longer influence the person in question. 

However, the friendships that have been terminated may not have be ``random.'' Relationships often end for a reason. If the reason for friendship termination is related to or is correlated with changes between $t_0$ and $t_1$ in the underlying trait we are examining, it will induce an association between $Y_{i,t_1}$ and $Y_{j,t_1}$ that is not captured by the lagged values of the variables in question.\footnote{The threat of non-random unfriending can be considered a specific example of the problem noted by Nickerson (\citealt{fowler11}) in his discussion of the value of conducting experiments on static networks: ``it is always possible that measurement of the network post-treatment could be correlated with the provision of the treatment. If treatments cause certain relationships to become more salient or networks to change composition, then many strategies for defining networks \ldots may cease to be equivalent for treatment and control groups.''} In the CF model, the coefficient for $Y_{j,t_1}$ is interpreted as a causal effect. As such, the association induced by homophily in the unfriending process could create the appearance of an influence effect even if none exists.\footnote{In principle, one might attempt to model the selection process by which friendships are maintained in order to recover the true value of the influence coefficient. However, it seems impossible to obtain data that is granular enough to separate stochastic changes in the trait of interest from $t_0$ to $t_1$ from peer effects. In the absence of such data, accurately modeling the friendship retention process requires knowing the value of $Y_{i,t_1}$ that would have been observed if no influence had taken place---an unobserved counterfactual.}

\section{Monte Carlo simulation procedure}

To determine the extent to which homophily in friendship retention might lead to spurious inferences under CF's model, we conduct Monte Carlo simulations in which we do not allow for social influence of alters on egos.\footnote{A broader question that we do not engage here is whether statistical models of such effects are formally identified. \citet{shalizi11} presents a graphical causal model arguing that such effects are generically unidentified in observational data for a person $i$ when some latent trait ``$X_i$ directly influences $Y_{i,t}$ for all $t$.'' In such cases, even controlling for $Y_{i,t-1}$ and $Y_{j,t-1}$ is not sufficient to identify the causal effect of a network tie $A_{i,j}$ on $Y_{i,t}$.} In reality, of course, such influence seems likely to be common. However, we follow standard practice in Monte Carlo evaluation of statistical models in assuming that the coefficient in question is zero and estimating the bias and coverage of the model. (The simulations reported in  \citet{fowlerjhe08} and \citet{fowler11} also include estimates of bias in the model where the true influence effect is equal to 0.) By working within this framework, which offers well-defined standards of model performance\footnote{By contrast, it is rare to test a model with a non-zero null hypothesis using frequentist statistics and it it not entirely clear how such a model should be evaluated.}, we can evaluate the the risk that homophily in friendship retention will lead researchers using the CF model to falsely reject the null (the standard inferential approach used in applications of the model).\footnote{These Type I errors could happen in a variety of ways. First, many researchers test hypotheses for which we have weak priors and a null hypothesis of zero may be a valid initial assumption. In such cases, it is important to know whether CF's model may generate spurious findings. Moreover, even if we have strong priors that social influence is taking place, we can never be sure that the expected effect is present in the data we are analyzing. It is important to know whether the CF approach could generate a spurious positive effect in those cases.}

In this section, we explain the Monte Carlo simulation procedure used in our analysis, which is adapted from code used in \citet{fowlerjhe08} and \citet{fowler11}. The R code used to generate our results will be posted on the \emph{Social Networks} website along with the electronic version of our article. 

The simulation proceeds as follows:
\begin{enumerate}
\item \label{init} A normally distributed trait $Y_{t_0}$ is randomly generated at time $t_0$ for a population of $n$=1000 actors where $Y_{t_0} \sim \mathrm{N}(50,10)$.\footnote{\citet{fowlerjhe08} draw their trait values from the empirical distribution of BMI in the Framingham data. Since we do not have access to those data, we use the initial distribution from the simulation in \citet{fowler11} instead.}
\item \label{difference} Differences in $Y$ are computed for all dyads of actors $i$ and $j$ in the same manner as CF:
\begin{equation*}
d_{i,j} = - |Y_{i,t_0} - Y_{j,t_0}|
\end{equation*}
The difference term is negatively valued so that dyads with similar traits have high values. 

\item \label{probit} Ties $A_{i,j}$ are created as a function of $d_{i,j}$ using a probit model based on a latent variable $A^*_{i,j}$. These ties are directed (i.e., $A_{i,j}$ does not imply $A_{j,i}$).\\

\begin{equation*}
A_{i,j} (t_0) = \left\{ 
\begin{array}{l l l}
  1 & \quad \text{if} & A^*_{i,j} > \epsilon_{i,j} \sim N(0,1) \\
  0 & \quad \text{if} & A^*_{i,j} \le \epsilon_{i,j} \sim N(0,1)\\
\end{array} \right.
\end{equation*}

where 

\begin{equation*}
A^*_{i,j} = \eta_0 + \eta_1 d_{i,j} 
\end{equation*}

$\eta_0$ represents the baseline propensity to form ties and $\eta_1$ represents the coefficient for homophily (positive values indicate higher levels of homophily). 

\item All actors receive a normally distributed, independent shock $u_i$ to their trait $Y_{t_0}$ where $u_i \sim \mathrm{N}(0,5)$.\footnote{\citet{fowlerjhe08} use shock values from the empirical distribution of changes in BMI in the Framingham data. Again, since we do not have  those data, we use the shock distribution from the simulation in \citet{fowler11} instead.}
\item In this step, CF assume that ego traits $Y_{t_1}$ are updated as a function of their previous trait value $Y_{t_0}$ and influence from their alters.\footnote{\label{cfeq}In CF's simulation, all egos' values of $Y$ are updated as a weighted average of their own current value of $Y_{t_0}+u_i$ and the average value of $Y_{t_0}+u_i$ for their alters:
\begin{equation*}
Y_{i,t_1} = (1-b_1)(Y_{i,t_0}+u_{i}) + b_1 \left(\frac{\sum_{j} A_{ij}(Y_{j,t_0}+u_j)}{\sum_{j} A_{ij}}\right)
\end{equation*}
where $b_1$ is represents the relative influence of alters on egos. Since we set peer influence to 0, our results are not sensitive to this assumption.} However, since we assume there is no peer influence, ego traits at $t_1$ are simply the sum of their previous value $Y_{t_0}$ and their shock $u_i$:
\begin{equation*}
Y_{i,t_1} = Y_{i,t_0}+u_{i}
\end{equation*}
\item \label{unfriendstage} All actors update their friendship ties $A_{i,j} (t_1)$ as in steps \ref{difference}--\ref{probit}. 
\item Following CF, we estimate a linear regression model using generalized estimating equations with an independent correlation structure for all dyads who are friends at $t_0$ and $t_1$:
\begin{equation}
Y_\mathrm{i,t_1}=\beta_0+Y_\mathrm{i,t_0}\beta_1  + Y_\mathrm{j,t_0} \beta_2 + Y_\mathrm{j,t_1} \beta_3 + \epsilon \mathrm{\, where \,} A_{i,j} (t_0, t_1) =1 
\end{equation}
\end{enumerate}

\begin{figure}[htbp]
\begin{center}
\caption{Sample network}
\label{samplenet}
\includegraphics[scale=.69]{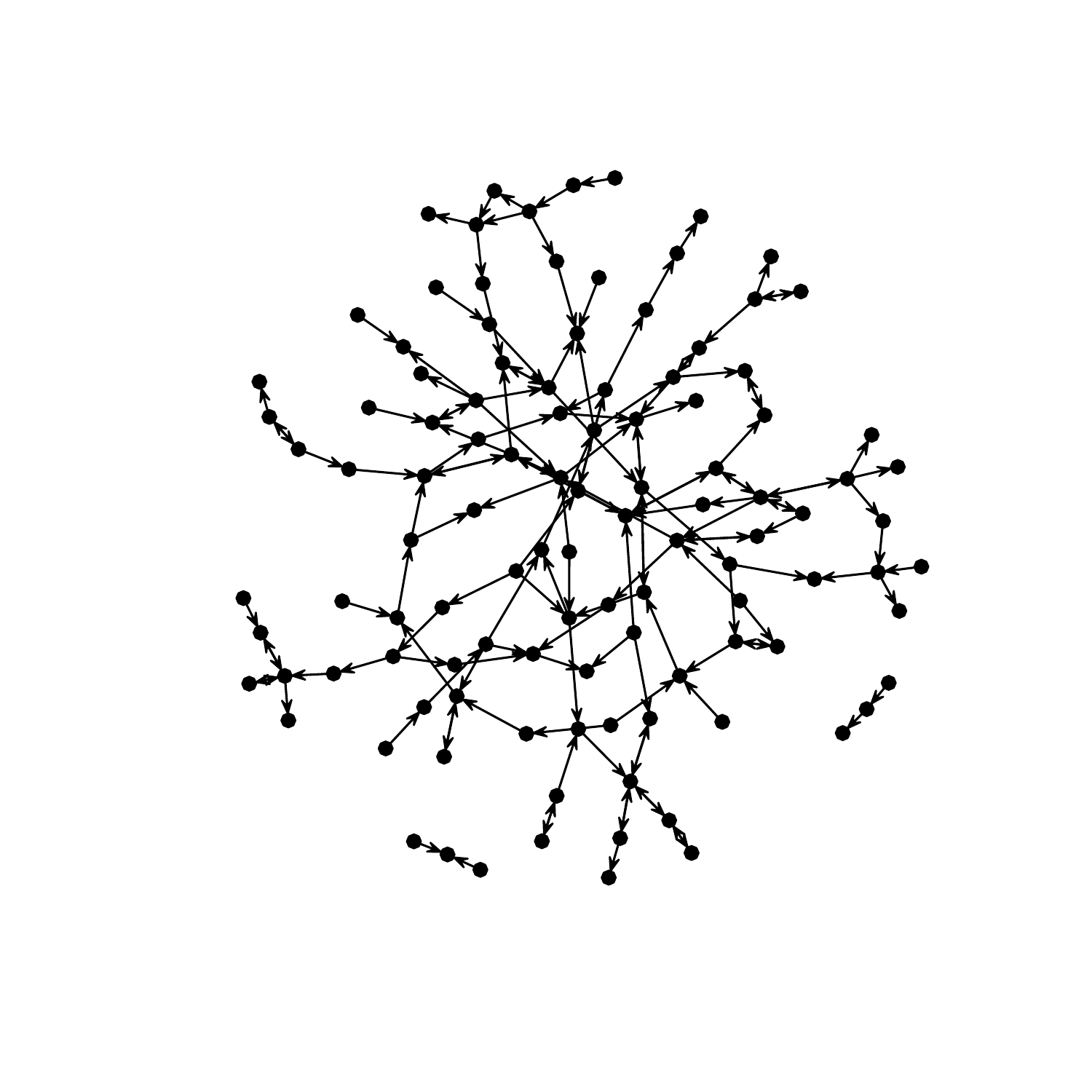}\\
\footnotesize{Sample network: $\eta_0$=1 in step \ref{unfriendstage}; $\eta_1$=0.05 in steps \ref{probit} and \ref{unfriendstage}; s.d.($u$)=5; $\eta_0$=-2.5; $n$=120 (isolates not displayed).}
\end{center}
\end{figure}

To illustrate the results, Figure \ref{samplenet} provides a sample network at the end of one simulation (here $n$=120 since networks of the size used in our simulations are too dense to parse visually). 

\begin{figure}[htbp]
\begin{center}
\caption{Illustration of one simulation}
\label{illustration}
\includegraphics[scale=.69]{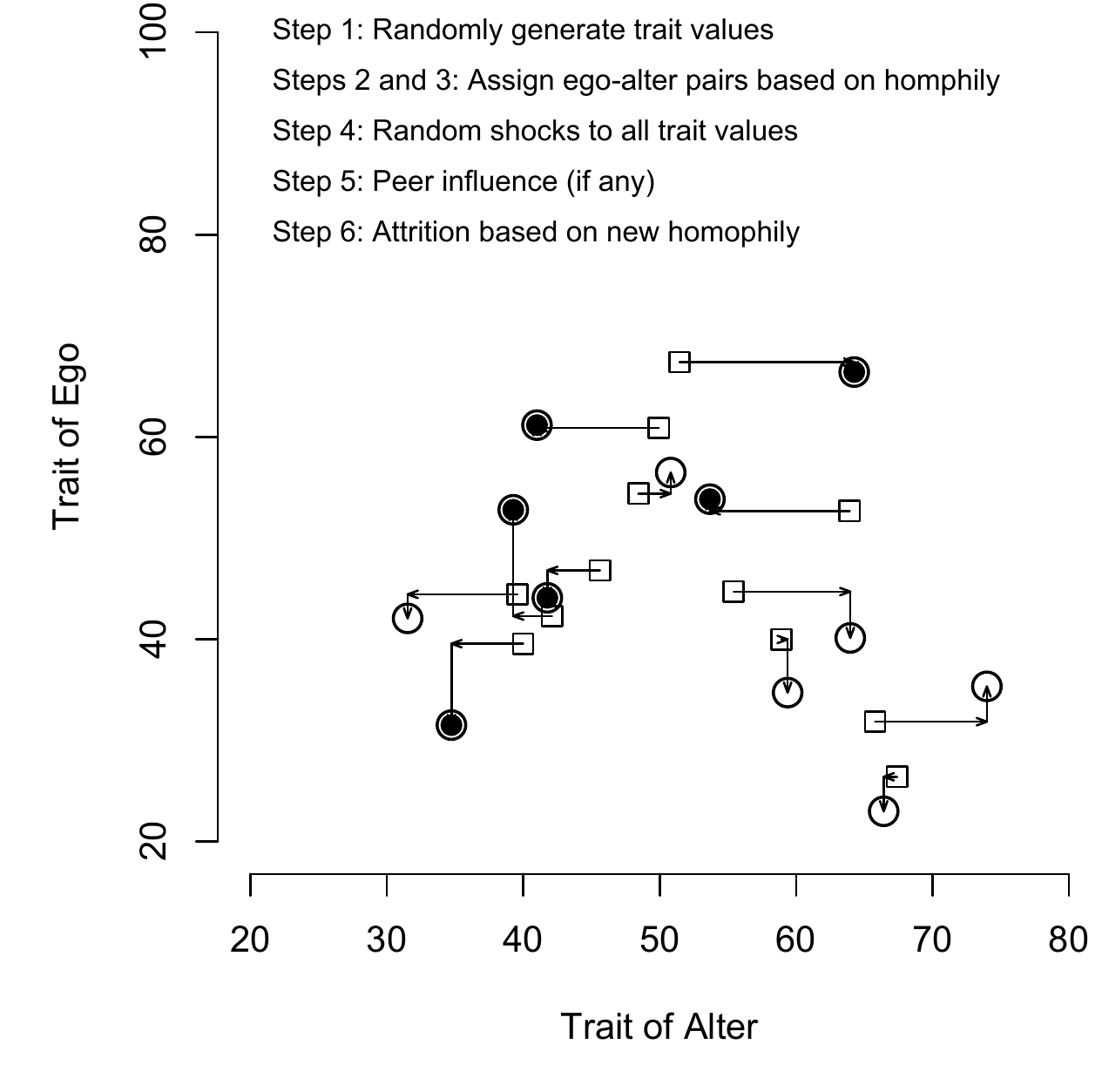}
\end{center}
\end{figure}

The simulation process is illustrated in Figure \ref{illustration}. Open squares represent values of $Y$ for ego-alter pairs after initial friendships have been formed. There is some correlation due to homophily. The actors then both experience shocks to their values of the trait, which are represented by arrows. The new values of the trait are indicated by the circles at the end of each path. The GEE should estimate the effect of the alter's shock on the ego controlling for the alter's previous $Y$ value. However, at the friendship retention stage, some of the pairs cease to be friends. Only those pairs indicated by the solid circles remain in the data; those that are open circles have ceased to be friends. 

This procedure above modifies CF's original approach in two key ways. The most important modification is step \ref{unfriendstage}, which repeats the friendship model from step \ref{probit}, allowing for friendships to end based on homophily after a shock to $Y$. This step is crucial in longitudinal network data as discussed above. While the shock $u$ to $Y_0$ is assumed to be randomly distributed, some people may cease to be friends as a result of (or for reasons that are correlated with) the shock that they received, inducing a correlation in $u$ for those dyads whose friendships persist. Our simulation is intended to test whether this correlation could appear to be a causal influence effect even if we control for lagged values of the trait.

The way we model friendship formation and dissolution also differs from CF's approach. We introduce a latent variable probit model where a tie exists if a deterministic component ($A^*_{i,j}=\eta_0 + \eta_1 d_{i,j}$) is greater than a stochastic error term ($\epsilon_{i,j} \sim \mathrm{N}(0,1)$). By contrast, CF generate a probability of a tie that is a weighted average of $Y_0$ and a random component and then conduct a random draw with that probability to determine whether a tie exists. This process combines two sources of random noise. The first is meant to model factors other than homophily in friendship choice, while the second models the inherently stochastic component of friendship formation. However, this partition is not readily interpretable---any unobservable influence on the outcome variable in a statistical model can reasonably be included in the stochastic component if it is not also systematically related to the independent variables. 

The result of this double-randomness is that CF's simulations do not generate sufficiently high correlations in the outcome variable among friends even when ties are formed on the basis of ``complete homophily'' \citep[1405]{fowlerjhe08}. When we replicated CF's simulations with homophily set to its maximum value, the mean correlation between ego and alter on the outcome variable was $0.12$ (2.5-97.5 percentile range: $0.05, 0.18$). In practice, social network datasets often display higher levels of correlation among friends. For instance, \citet{halliday09} find a BMI correlation of $0.19$ among adolescent peers in the Add Health data after accounting for school fixed effects. The correlation in reported vote choice between respondents to the 2000 American National Election Study and their friends ranged from $0.43$ to $0.57$ (results available upon request). Even accounting for the effects of projection (i.e., respondents falsely perceiving that their friends agree with them), these results suggest that simulations should consider higher levels of homophily. While some of these correlations may be due to contagion, we cannot assume such an effect. As such, we designed our simulations to cover a wide range of homophily on the trait of interest. 


\section{Monte Carlo results}\label{results}

Following standard procedure in Monte Carlo evaluations of statistical models, we set the true contagion effect (i.e., the parameter $b_1$) to 0 and estimate mean bias and coverage levels for the CF model. In our simulations, we set $\eta_0$ to -2.5 at the friendship formation stage in order to generate realistic numbers of friendships at $t_0$.\footnote{CF's Framingham subjects typically only name one friend due to the nature of the instrument used. However, this structure is unusual and we do not mimic it here. (See \citealt{thomasnd} for a related discussion of how binary networks with censored outdegree information may generate inflated social influence effects.)} We then vary $\eta_1$ at both stages to generate realistic levels of homophily in both friendship formation and retention and also vary $\eta_0$ at the friendship retention stage to consider different friendship attrition rates.\footnote{In practice, it is possible that homophily in friendship formation and retention could generate feedback effects in a more elaborate multi-stage simulation (as, for instance, new, more similar acquaintances displace old, less similar friends), but we do not attempt to model the complexities of those potential dynamics here, especially given our lack of substantive knowledge about how such a process would operate.}
\begin{itemize}
\item Homophily in friendship formation: We vary the homophily parameter $\eta_1$ in step \ref{probit} to cover five possible levels: $0.0$, $0.0125$, $0.025$, $0.0375$, and $0.05$.
\item Homophily in friendship retention: We vary the homophily parameter $\eta_1$ in step \ref{unfriendstage} to cover three levels: $0.0$, $0.025$, and $0.05$.
\item Levels of friendship retention: We vary $\eta_0$ in step \ref{unfriendstage}, considering values of $0.0$, $0.5$, and $1.0$ to represent realistic variation in attrition rates between $t_0$ and $t_1$. We also include the value of $1.85$ to cover the maximum FHS friendship retention rate of 0.96 (see footnote \ref{point96fn} above).\footnote{Additional simulations in which we also vary the standard deviation of the shock $u$ to $Y_0$ generate similar results. They are thus omitted but available on request.}
\end{itemize}

These values correspond to realistic levels of homophily and friendship retention. For instance, \cite{omalleynd} use FHS data to test whether they observe homophily in friendship formation and dissolution using BMI and related measures. Their hierarchical logit models find no effect for homophily at the formation stage, which would correspond to a $\eta_1$ value of $0.0$ in step \ref{probit}, but find a statistically significant and substantively large coefficient on BMI for friendship retention. While we cannot directly compare coefficient values given differences in data and estimation techniques, the value of $0.05$ for $\eta_1$ in step \ref{unfriendstage} appears to be an appropriate comparison.

Complete results from the Monte Carlo simulations, which were performed 1,000 times for each unique combination of model parameters, are presented in Table \ref{resultssummary} at the end of the document. The first result of note is that our simulations cover realistic ranges of both ego-alter trait correlation and friendship attrition. As we discuss above, observed ego-alter trait correlations can be quite high. In the simulations, these correlations range from approximately $0.0$ to $0.6$ at $t_0$ and $0.0$ to $0.4$ at $t_1$. Similarly, the current simulation yields friendship retention rates of approximately 50\% in the high attrition case (retention constant $\eta_0=0$), 69\% in the moderate attrition case ($\eta_0=0.5$), 84\% in the low attrition case ($\eta_0=1.00$), and 97\% in the very low attrition case ($\eta_0=1.85$).

\begin{sidewaysfigure}
\begin{center}
\caption{Estimated coverage rates for CF model in Monte Carlo simulations}
\label{coverage}
\includegraphics[scale=.65]{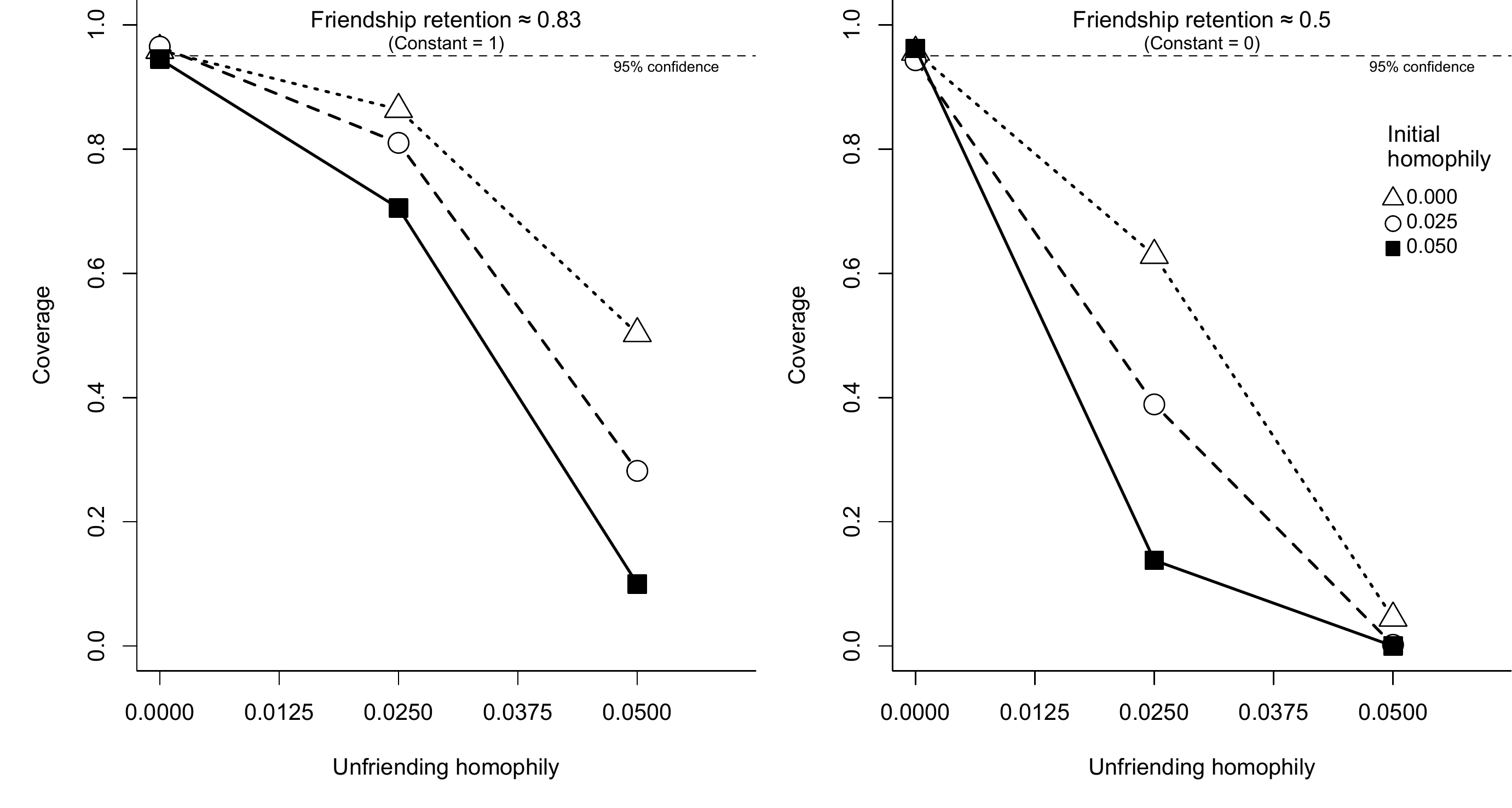}
\end{center}
\end{sidewaysfigure}

\begin{sidewaysfigure}
\begin{center}
\caption{Estimated bias for CF model in Monte Carlo simulations}
\label{bias}
\includegraphics[scale=.65]{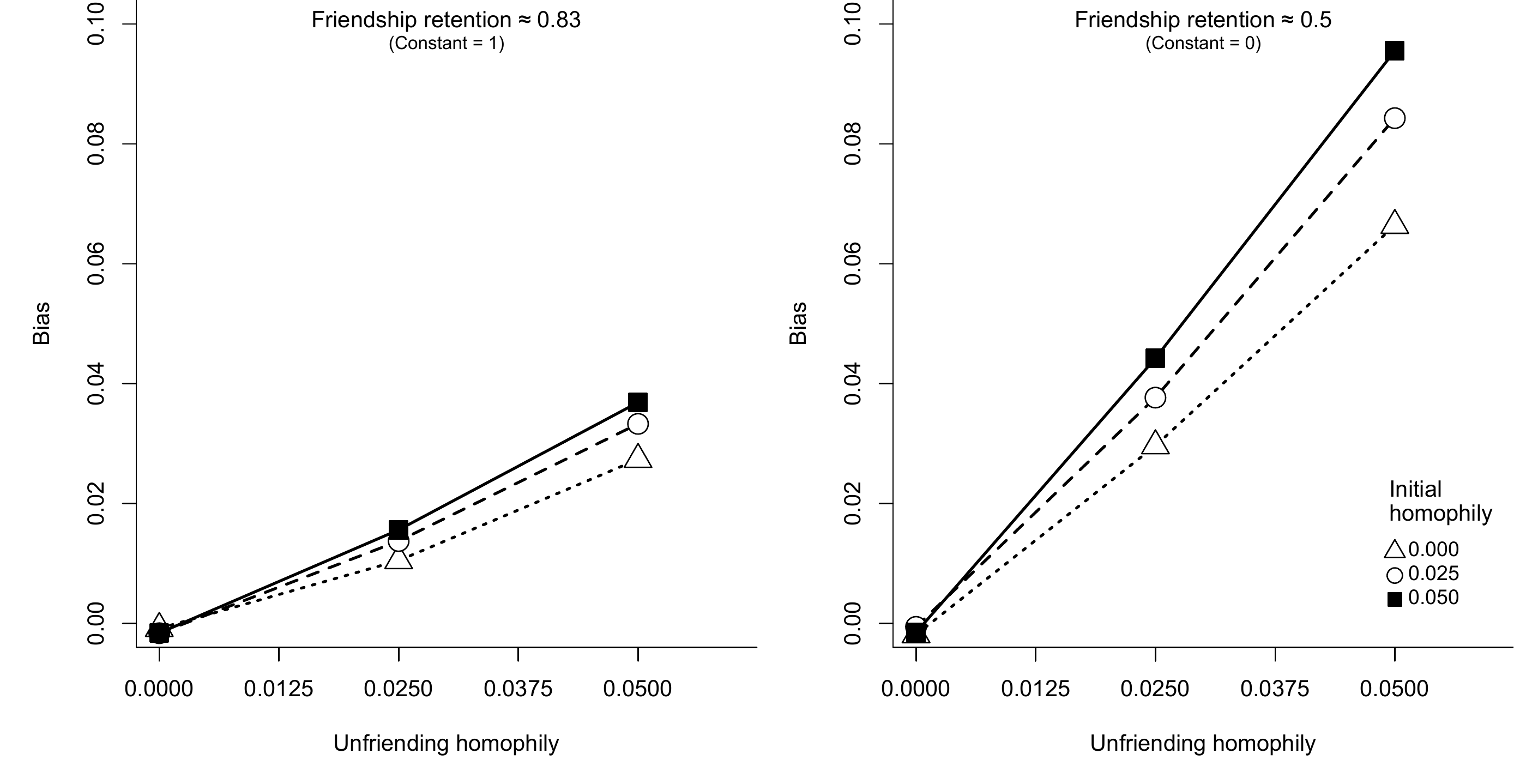}
\end{center}
\end{sidewaysfigure}

Figures \ref{coverage} and \ref{bias} plot how well the GEE estimate of equation 2 performed for varying values of the homophily coefficients when the constant $\eta_0=1$ and $\eta_0=0$ at the friendship retention stage (those for $\eta_0=0.5$ and $\eta_0=1.85$ are not plotted but are available in Table 1). These values approximately correspond to attrition rates in FHS (18\%) and Add Health (50\%), respectively. For visual clarity, the figures include bias and coverage levels for $\eta_1$ at the friendship formation stage of 0, 0.025, and 0.05. (When $\eta_1$ equals $0.0125$ and $ 0.0375$ at the friendship formation stage, the results fall between the three lines in Figures \ref{coverage} and \ref{bias}. They are thus omitted from the figures but described in Table 1.)

First, Figure \ref{coverage} presents the probability that the estimated 95\% confidence interval covers the true contagion parameter of 0.
When homophily in friendship retention is 0, the confidence intervals for the CF models accurately bracket the true value approximately 95 percent of the time. However, as homophily in friendship retention increases, coverage rates decline dramatically. For instance, when initial homophily is 0 and friendship retention is high (the open triangles in the left panel of Figure \ref{coverage}), coverage falls to approximately 50 percent as homophily in friendship retention increases. In the most extreme cases, the confidence interval almost never includes the true value of the influence coefficient. Even when friendship attrition is very low (approximately 4\%), Table 1 indicates that the CF model still has some coverage problems (decreasing as low as 81 percent) when homophily in friendship retention is high (see lines 46 to 60 of Table 1). In additional simulations, we find that coverage problems worsen further as sample size increases, thereby increasing the likelihood that the CF model will falsely reject the null hypothesis that the influence effect is 0 (in practice, CF's data frequently include far more than 1000 observations). 

As expected, coverage degrades because the model displays an upward bias, as is evident in Figure \ref{bias}, which presents the mean value of the estimated peer effect (which has been set to 0 in the simulations). 
When unfriending is not affected by homophily, the estimator is unbiased. But as homophily in friendship retention increases, a correlation emerges in changes in the trait of interest between egos and alters, which the model interprets as evidence of social influence. As a result, estimated bias levels increase substantially---up to 0.10 in the worst case. Again, even in the left panel where friendship attrition is not large, the bias is substantial if friendship attrition is closely related to homophily. 

Is the level of bias described above meaningful? As a point of comparison, we note that estimated peer effect coefficients for continuous variables in the literature are often in the range described in Table \ref{resultssummary} and Figure \ref{bias}. For example, the coefficient on alter's current BMI is 0.053 in the FHS data (SE=0.018) and 0.033 in Add Health data (SE=0.014) \citep{fowlerjhe08}.\footnote{It would be worthwhile to repeat this exercise with a binary variable as in the CF studies of smoking or depression.} Of course, our simulations cannot prove that the results in any given study are spurious, but they do suggest that the risk of falsely rejecting the null hypothesis is high for the CF model when homophily in friendship retention is present. 

\section{Conclusion}

In this paper, we have argued that the ``unfriending'' problem complicates efforts to estimate causal influence effects in longitudinal social network data. While previous studies have demonstrated how homophily in friendship formation can confound estimates of peer effects, we are the first to demonstrate that homophily in friendship \emph{retention} can create a similar problem by inducing an association between random shocks to an outcome variable for dyads that remain friends after the shock. We provide evidence for this hypothesis using an adaptation of Christakis and Fowler's Monte Carlo simulation. Our simulations show that, when friendship attrition is present, the CF model suffers from serious bias and coverage problems as homophily in friendship retention increases. This association is large enough that, under certain parameter values, it could account for most or even all of the observed associations in published estimates. 

These results suggest that caution is required in interpreting findings of peer effects that rely on maintained ties over time. Though this paper focuses on friendship network data, the logic of  the ``unfriending'' problem applies to estimates of peer effects in any longitudinal network data in which homophily plays a significant role in tie dissolution. For instance, \citet{grannis10} discusses how inferences about network structure change if one allows for tie dissolution over time in the Ph.D. exchange network among sociology graduate programs. If there is significant homophily in tie retention in this network (for instance, by methodological or substantive focus), one might obtain misleading estimates of ``influence'' effects among graduate programs. 

Going forward, more research is clearly needed on models of peer influence in observational data. For instance, the actor-based model of \citet{Steglichetal2010} attempts to account for many of the concerns described above. However, our simulation results suggest that it is essential to test the properties of such models using simulations in which the true parameter values are known. 

In addition, more research is needed on how to \emph{assess} models of peer influence. While the parsimonious approach presented here has limitations, it provides a flexible framework for more complex simulations of friendship ties and social influence. Many desirable modifications are possible, including incorporating more than two time periods; allowing peer effects to vary by friendship duration; estimating the effect of latent traits; including environmental confounders; or allowing for correlations in shocks among peers. As research proceeds on models of network formation (e.g., \citealt{christakisnd}), it may soon also be possible to simulate random networks that more closely mirror important features of human networks such as clustering, mutuality, and transitivity. As our theoretical understanding of networks improves, so too will our ability to test statistical models of social influence in observational data.

\newpage
\singlespacing
\normalsize
\bibliographystyle{apsr1.bst}
\bibliography{netcaus.bib}

\begin{landscape}
\begin{longtable}{ccccccccccc}

\caption{Simulation results}\\
\hline
& Retention & Formation &  Retention  &       &                 &          Ego-alter          &   Ego-alter               & Friends/ & Friends/ & Friendship\\
& constant  & homophily &  homophily  &       & Coverage & correlation & correlation & person & person & retention \\
& $\eta_0$ ($t_1$) & $\eta_1$ ($t_0$)  &  $\eta_1$  ($t_1$)   & Bias & probability & ($t_0)$        & ($t_1$)        & ($t_0$)     & ($t_1$) & rate\\
\hline
\endfirsthead
\caption[]{Simulation results (continued)}\\

\hline
& Retention & Formation &  Retention  &       &                 &          Ego-alter          &   Ego-alter               & Friends/ & Friends/ & Friendship\\
& constant  & homophily &  homophily  &       & Coverage & correlation & correlation & person & person & retention \\
& $\eta_0$ ($t_1$) & $\eta_1$ ($t_0$)  &  $\eta_1$  ($t_1$)   & Bias & probability & ($t_0)$        & ($t_1$)        & ($t_0$)     & ($t_1$) & rate\\
\hline
\endhead

\hline
\multicolumn{11}{l}{$\eta_0=$$-2.75$ at $t_0$; true contagion effect $b_1$=$0$; $u \sim \mathrm{N}(0,5)$; $n$=$1000$; 1000 simulations per row (continued on next page)}\\

\endfoot

\hline
\multicolumn{11}{l}{$\eta_0=$$-2.75$ at $t_0$; true contagion effect $b_1$=$0$; $u \sim \mathrm{N}(0,5)$; $n$=$1000$; 1000 simulations per row}\\
\endlastfoot
1 & 0.00 & 0.0000 & 0.000 & -0.00 & 0.96 & -0.00 & -0.00 & 6.2 & 3.1 & 0.50 \\ 
  2 & 0.00 & 0.0000 & 0.025 & 0.03 & 0.63 & -0.00 & -0.00 & 6.2 & 3.1 & 0.50 \\ 
  3 & 0.00 & 0.0000 & 0.050 & 0.07 & 0.05 & -0.00 & -0.00 & 6.2 & 3.1 & 0.51 \\ 
  4 & 0.00 & 0.0125 & 0.000 & -0.00 & 0.96 & 0.19 & 0.15 & 6.4 & 3.2 & 0.50 \\ 
  5 & 0.00 & 0.0125 & 0.025 & 0.03 & 0.54 & 0.19 & 0.15 & 6.4 & 3.2 & 0.50 \\ 
  6 & 0.00 & 0.0125 & 0.050 & 0.08 & 0.01 & 0.19 & 0.15 & 6.4 & 3.2 & 0.50 \\ 
  7 & 0.00 & 0.0250 & 0.000 & -0.00 & 0.94 & 0.36 & 0.27 & 7.1 & 3.5 & 0.50 \\ 
  8 & 0.00 & 0.0250 & 0.025 & 0.04 & 0.39 & 0.36 & 0.27 & 7.1 & 3.5 & 0.50 \\ 
  9 & 0.00 & 0.0250 & 0.050 & 0.08 & 0.00 & 0.36 & 0.27 & 7.1 & 3.6 & 0.50 \\ 
  10 & 0.00 & 0.0375 & 0.000 & -0.00 & 0.94 & 0.49 & 0.36 & 8.1 & 4.1 & 0.50 \\ 
  11 & 0.00 & 0.0375 & 0.025 & 0.04 & 0.26 & 0.49 & 0.36 & 8.1 & 4.1 & 0.50 \\ 
  12 & 0.00 & 0.0375 & 0.050 & 0.09 & 0.00 & 0.49 & 0.36 & 8.1 & 4.1 & 0.50 \\ 
  13 & 0.00 & 0.0500 & 0.000 & -0.00 & 0.96 & 0.59 & 0.42 & 9.6 & 4.8 & 0.50 \\ 
  14 & 0.00 & 0.0500 & 0.025 & 0.04 & 0.14 & 0.59 & 0.42 & 9.6 & 4.8 & 0.50 \\ 
  15 & 0.00 & 0.0500 & 0.050 & 0.10 & 0.00 & 0.59 & 0.42 & 9.6 & 4.8 & 0.50 \\ 
  16 & 0.50 & 0.0000 & 0.000 & -0.00 & 0.94 & -0.00 & -0.00 & 6.2 & 4.3 & 0.69 \\ 
  17 & 0.50 & 0.0000 & 0.025 & 0.02 & 0.77 & -0.00 & -0.00 & 6.2 & 4.3 & 0.69 \\ 
  18 & 0.50 & 0.0000 & 0.050 & 0.05 & 0.16 & -0.00 & -0.00 & 6.2 & 4.2 & 0.68 \\ 
  19 & 0.50 & 0.0125 & 0.000 & -0.00 & 0.94 & 0.19 & 0.15 & 6.4 & 4.4 & 0.69 \\ 
  20 & 0.50 & 0.0125 & 0.025 & 0.02 & 0.69 & 0.19 & 0.15 & 6.4 & 4.4 & 0.69 \\ 
  21 & 0.50 & 0.0125 & 0.050 & 0.05 & 0.10 & 0.19 & 0.15 & 6.4 & 4.4 & 0.68 \\ 
  22 & 0.50 & 0.0250 & 0.000 & -0.00 & 0.95 & 0.36 & 0.27 & 7.1 & 4.9 & 0.69 \\ 
  23 & 0.50 & 0.0250 & 0.025 & 0.02 & 0.63 & 0.36 & 0.27 & 7.1 & 4.9 & 0.69 \\ 
  24 & 0.50 & 0.0250 & 0.050 & 0.06 & 0.03 & 0.36 & 0.27 & 7.1 & 4.8 & 0.68 \\ 
  25 & 0.50 & 0.0375 & 0.000 & -0.00 & 0.95 & 0.49 & 0.36 & 8.1 & 5.6 & 0.69 \\ 
  26 & 0.50 & 0.0375 & 0.025 & 0.03 & 0.50 & 0.49 & 0.36 & 8.1 & 5.6 & 0.69 \\ 
  27 & 0.50 & 0.0375 & 0.050 & 0.06 & 0.01 & 0.49 & 0.36 & 8.1 & 5.6 & 0.68 \\ 
  28 & 0.50 & 0.0500 & 0.000 & -0.00 & 0.95 & 0.59 & 0.42 & 9.6 & 6.6 & 0.69 \\ 
  29 & 0.50 & 0.0500 & 0.025 & 0.03 & 0.40 & 0.59 & 0.42 & 9.6 & 6.6 & 0.69 \\ 
  30 & 0.50 & 0.0500 & 0.050 & 0.06 & 0.00 & 0.59 & 0.42 & 9.6 & 6.6 & 0.68 \\ 
  31 & 1.00 & 0.0000 & 0.000 & -0.00 & 0.96 & -0.00 & -0.00 & 6.2 & 5.2 & 0.84 \\ 
  32 & 1.00 & 0.0000 & 0.025 & 0.01 & 0.86 & -0.00 & -0.00 & 6.2 & 5.2 & 0.83 \\ 
  33 & 1.00 & 0.0000 & 0.050 & 0.03 & 0.50 & -0.00 & -0.00 & 6.2 & 5.1 & 0.82 \\ 
  34 & 1.00 & 0.0125 & 0.000 & -0.00 & 0.95 & 0.19 & 0.15 & 6.4 & 5.4 & 0.84 \\ 
  35 & 1.00 & 0.0125 & 0.025 & 0.01 & 0.87 & 0.19 & 0.15 & 6.4 & 5.4 & 0.84 \\ 
  36 & 1.00 & 0.0125 & 0.050 & 0.03 & 0.37 & 0.19 & 0.15 & 6.4 & 5.3 & 0.82 \\ 
  37 & 1.00 & 0.0250 & 0.000 & -0.00 & 0.96 & 0.36 & 0.27 & 7.1 & 6.0 & 0.84 \\ 
  38 & 1.00 & 0.0250 & 0.025 & 0.01 & 0.81 & 0.36 & 0.27 & 7.1 & 5.9 & 0.84 \\ 
  39 & 1.00 & 0.0250 & 0.050 & 0.03 & 0.28 & 0.36 & 0.27 & 7.1 & 5.9 & 0.83 \\ 
  40 & 1.00 & 0.0375 & 0.000 & -0.00 & 0.94 & 0.49 & 0.36 & 8.1 & 6.9 & 0.84 \\ 
  41 & 1.00 & 0.0375 & 0.025 & 0.01 & 0.78 & 0.49 & 0.36 & 8.1 & 6.8 & 0.84 \\ 
  42 & 1.00 & 0.0375 & 0.050 & 0.04 & 0.17 & 0.49 & 0.36 & 8.1 & 6.7 & 0.83 \\ 
  43 & 1.00 & 0.0500 & 0.000 & -0.00 & 0.94 & 0.59 & 0.42 & 9.6 & 8.1 & 0.84 \\ 
  44 & 1.00 & 0.0500 & 0.025 & 0.02 & 0.71 & 0.59 & 0.42 & 9.6 & 8.1 & 0.84 \\ 
  45 & 1.00 & 0.0500 & 0.050 & 0.04 & 0.10 & 0.59 & 0.42 & 9.6 & 8.0 & 0.83 \\ 
  46 & 1.85 & 0.0000 & 0.000 & -0.00 & 0.95 & -0.00 & -0.00 & 6.2 & 6.0 & 0.97 \\ 
  47 & 1.85 & 0.0000 & 0.025 & 0.00 & 0.97 & -0.00 & -0.00 & 6.2 & 6.0 & 0.96 \\ 
  48 & 1.85 & 0.0000 & 0.050 & 0.01 & 0.88 & -0.00 & -0.00 & 6.2 & 5.9 & 0.95 \\ 
  49 & 1.85 & 0.0125 & 0.000 & -0.00 & 0.94 & 0.19 & 0.15 & 6.4 & 6.2 & 0.97 \\ 
  50 & 1.85 & 0.0125 & 0.025 & 0.00 & 0.96 & 0.19 & 0.15 & 6.4 & 6.2 & 0.96 \\ 
  51 & 1.85 & 0.0125 & 0.050 & 0.01 & 0.86 & 0.19 & 0.15 & 6.4 & 6.1 & 0.95 \\ 
  52 & 1.85 & 0.0250 & 0.000 & -0.00 & 0.94 & 0.36 & 0.27 & 7.1 & 6.9 & 0.97 \\ 
  53 & 1.85 & 0.0250 & 0.025 & 0.00 & 0.94 & 0.36 & 0.27 & 7.1 & 6.8 & 0.97 \\ 
  54 & 1.85 & 0.0250 & 0.050 & 0.01 & 0.88 & 0.36 & 0.27 & 7.1 & 6.8 & 0.96 \\ 
  55 & 1.85 & 0.0375 & 0.000 & -0.00 & 0.95 & 0.49 & 0.36 & 8.1 & 7.9 & 0.97 \\ 
  56 & 1.85 & 0.0375 & 0.025 & 0.00 & 0.93 & 0.49 & 0.36 & 8.1 & 7.9 & 0.97 \\ 
  57 & 1.85 & 0.0375 & 0.050 & 0.01 & 0.84 & 0.49 & 0.36 & 8.1 & 7.8 & 0.96 \\ 
  58 & 1.85 & 0.0500 & 0.000 & -0.00 & 0.95 & 0.59 & 0.42 & 9.6 & 9.3 & 0.97 \\ 
  59 & 1.85 & 0.0500 & 0.025 & 0.00 & 0.93 & 0.59 & 0.42 & 9.6 & 9.3 & 0.97 \\ 
  60 & 1.85 & 0.0500 & 0.050 & 0.01 & 0.81 & 0.59 & 0.42 & 9.6 & 9.2 & 0.96 \\

\label{resultssummary}
\end{longtable}
\end{landscape}

\end{document}